\documentclass[12pt,preprint]{aastex}

\slugcomment{}

\shorttitle{3D Global Corona near Solar Maximum}
\shortauthors{M. Kramar {\it et al.}}

\usepackage{hyperref}
\usepackage{graphicx}                    
\usepackage{color}                       
\usepackage{url}                         
\usepackage{ulem}

\newcommand{\svmdkl}{    Soviet Math. Dokl.}
\newcommand{\bsawiss}{    {\it Ber. S\"{a}chs. Akad. Wiss.}}

\usepackage{bm}
\renewcommand{\vec}[1]{\bm{#1}}

\begin{document}

\title{3D Global Coronal Density Structure and Associated Magnetic Field  near Solar Maximum}

\author{M. Kramar\altaffilmark{1}, V. Airapetian\altaffilmark{2,3}, H. Lin\altaffilmark{4}}

\affil{\altaffilmark{1}Physics Department, The Catholic University of America, 
620 Michigan Ave NE, Washington, DC 20064}

\affil{\altaffilmark{2}NASA-GSFC, Code 671, Greenbelt, MD 20771, USA}

\affil{\altaffilmark{3}Department of Physics and Astronomy, George Mason University, Fairfax, VA 22030, USA}

\affil{\altaffilmark{4}Institute for Astronomy, University of Hawaii at Manoa, 34 Ohia Ku Street, Pukalani, Maui, HI 96768, USA}

\altaffiltext{1}{e-mail: \url{kramar@cua.edu}}


\begin{abstract}
Measurement of the coronal magnetic field is a crucial ingredient in understanding 
the nature of solar coronal  dynamic phenomena at all scales. 
We employ STEREO/COR1 data obtained near maximum of solar activity in December 2012 
(Carrington rotation, CR 2131) to 
retrieve and analyze the three-dimensional (3D) coronal electron density in the range of heights 
from $1.5$ to $4\ \mathrm{R}_\odot$ using a tomography method 
and qualitatively deduce structures of the coronal magnetic field. 
The 3D electron density analysis is complemented by the 3D STEREO/EUVI emissivity in 195 \AA \ band obtained 
by tomography for the same CR period. 
We find that the magnetic field configuration during CR 2131 has a tendency to become radially open at heliocentric distances below $\sim 2.5 \ \mathrm{R}_\odot$. 
We compared the reconstructed 3D coronal structures over the CR near the solar maximum to the one 
at deep solar minimum. Results of our 3D density reconstruction will help to constrain solar coronal field models and test the accuracy of 
the magnetic field approximations for coronal modeling. 
\end{abstract}


\section{Introduction}

Solar coronal magnetic field is a major source of dynamics and thermodynamics of global coronal corona and transient coronal events including the coronal heating, solar flares, coronal mass ejections and the solar wind. Its dynamics affects space weather processes that may impact Earth's magnetosphere and atmosphere and affect life on our planet.  Thus, the knowledge of coronal magnetic field strength and topology represents one of major  goals of the solar physics. Currently, coronal magnetic fields cannot directly measured. The major techniques that are currently used to derive the global magnetic structures of the solar corona represent indirect methods including potential field source surface (PFSS) models, non-linear force-free field (NLFFF) models and multi-dimensional magnetohydrodynamic (MHD) models of the global solar corona. These methods use solar photospheric scalar or vector magnetic field as input directly derived from photospheric magnetograms. However, as we discussed in \cite{Kramar_2014}, all these methods cannot adequately describe the dynamics of magnetic fields observed in the solar corona driven by current carrying structures. Moreover, they do not provide accurate information about the coronal thermodynamics, and thus cannot be used for modeling the coronal emission measures derived from extreme ultraviolet (EUV) observations. 
	The MHD modeling of the solar corona provides a self-consistent time-dependent treatment of the plasma pressure, gravitational and magnetic forces, but are limited by approximations used for describing the coronal heating, and the uncertainties in the boundary conditions that are deduced from synoptic data. Thus, direct measurements of the coronal magnetic field remains one of the most reliable and challenging ways for characterization of solar coronal processes.

Direct measurement of the coronal magnetic field is the most challenging problem in observational solar physics. 
A major progress here was reached with the deployment of the Coronal Multichannel Polarimeter (CoMP) \citep{Tomczyk_2007,Tomczyk_2008}. 
In order to interpret such type of data, 
a vector tomography method has been developed for 3D reconstruction of the global coronal magnetic field \citep{Kramar_2006, Kramar_2013}, 
and recently, first 3D reconstruction of the global coronal magnetic field 
has been performed based on the CoMP data \citep{Kramar_2016}.

Another, thought implicit but also based on coronal observations, 
way to reconstruct some global coronal magnetic field structures 
was investigated in \citep{Kramar_2014}. 
There we applied the tomography method that employs STEREO data to reconstruct 3D density of global solar corona and related it to the coronal magnetic field structure. 
This method was applied to characterize the solar corona over the Carrington rotation 2066 occurred during deep solar minimum for the range of coronal heights from $1.5$ to $4 \ \mathrm{R}_\odot$ and provided description of the open/close magnetic field boundaries in the solar corona. 

In this paper, we apply this methodology to study global solar coronal structures for the period of over the half of Carrington rotation 2131, which represents the case of solar maximum. Specifically, we use the STEREO/COR1 coronagraph data for half a solar rotation period 
during CR 2131 as input for the tomographic reconstruction of the 3D coronal electron density and the configuation of magnetic field associated with them.  Our results are complemented by the 3D emissivity obtained by tomography method applied for the the STEREO  {\it Extreme Ultraviolet Imager} (EUVI) data in the 195 \AA \ band.  We also compared the reconstructed 3D coronal structures with those implied by the PFSS model. In Section 4, we present the comparison of the 3D coronal structure near the solar maximum to those at deep solar minimum (CR 2066).

\section{Methods}

The corona is optically thin at visible and EUV wavelengths dominated by lines of ionized iron and calcium. 
Thus, the observed visible and EUV coronal flux represents an integral of the emissivity along any line of sight (LOS).
This makes it impossible to reconstruct the spatial distribution of the
emissivity from a single (in a geometric sense) measurement or projection.
The solution space is reduced if we have measurements from many different viewpoints. 
The three-dimensional (3D) reconstruction of optically thin emission over depths can be performed by using tomography methods that employ the 2D observations of an object from different view angles. 
The possibility of reconstructing a function from its projections was first studied by \cite{Radon_1917}. 
Several decades later, this purely mathematical method formed the basis for 
the X-ray computer tomography that produces cross-sectional slices of human bodies \citep{Cormack_1963, Cormack_1964} and \citep{Hounsfield_1972}. 
Today, tomography is used in many fields of science including medical research, material structure testing, geophysics, heliophysics and astrophysics \citep{Boffin_2001}. 
In astrophysical applications the input data can suffer from noise and data incompleteness. While the application of tomography to astrophysical objects is limited by the noisy input data and its incompleteness, the regularization method allows solar coronal tomography to produce reliable 3D reconstructions of emissivity and plasma density
\citep{Tikhonov_1963, Frazin_2002, Kramar_2006, Kramar_2009}.

\subsection{Tomography Based on White-Light STEREO/COR1 Data}
\label{Sect_Tomo_COR1}

The key input for tomographic reconstruction of the plasma density of extended solar coronal structures requires observational data from more than two vantage points. Tomography applications for coronal studies typically assume a rigid rotation of the coronal density structures. For reliable density reconstruction, the algorithm utilizes coronagraph data for half a solar rotation as input if observed from a single spacecraft and requires  coronal structures to remain stable over their observation periods 
\citep{Davila_1994, Zidowitz_1999, Frazin_2005, Kramar_2009}. 
However, depending on the positions of a coronal structure relative to the spacecraft during the observation period, 
the stationarity assumption for that structure can be reduced to about a week \citep{Kramar_2011}.

For our density reconstructions we used the polarized brightness (pB) intensity images from the COR1 instrument 
onboard the STEREO-B spacecraft 
taken 28 images per half a solar rotation as input for the tomographic inversion. 
We limited here the data input for the tomography based on COR1 data to the STEREO-B spacecraft 
because COR1-B had lower levels of stray light during CR 2066 than COR1-A.

In the STEREO/COR1-B field of view (below $\approx 4\ \mathrm{R}_\odot$), the white-light pB coronal emission is dominated by 
scattering sunlight on the free electron in the corona 
\citep{Blackwell_1966a, Blackwell_1966b, Moran_2006, Frazin_2007}. 
The intensity of the pB-signal as a fraction of the mean solar brightness is given as 
\begin{equation}
I_\textrm{\small{pB}}(\vec{\hat{e}}_\textrm{\scriptsize{LOS}},\vec{\rho})=
\int \limits_\textrm{LOS} K(\vec{r})  N_e(\vec{r}) \mathrm{d} \ell ,
\label{ThompsonScat_pB}
\end{equation}
where $N_e$ is the electron density, $\vec{\rho}$ is a vector in plane-of-sky (POS) from the Sun center to the LOS 
and perpendicular to LOS, $\ell$ is length along the LOS, 
$\vec{\hat{e}}_\textrm{\scriptsize{LOS}}$ is the unit vector along the LOS, 
and $\vec{r}$ is the radius-vector. 
The kernel function [$K$] is defined by the Thompson scattering effect 
\citep{van_de_Hulst_1950BAN11_135V, Billings_1966, Quemerais_2002}: 
\begin{equation}
K=\frac{\pi \sigma}{2\left (1-\frac{u}{3} \right )}
\left [ (1-u)A(r)+uB(r) \right ] \frac{\rho^2}{r^2}
\label{KernelFunction_pB}
\end{equation}
where the expressions for $A(r)$ and $B(r)$ are the same as those given by \cite{Quemerais_2002}, 
$\sigma=7.95\times 10^{-26}\ \textrm{cm}^2$ is the Thompson scattering cross-section for a single electron, 
and the linear limb-darkening coefficient [$u$] is set to $0.6$ in the present calculations.

Because COR1 views the corona close to the limb, the instrument has a significant amount of scattered light, 
which must be subtracted from the image 
prior to be applied in the reconstruction method. 
Proper removal of instrumental scattered light is essential for coronal reconstruction.  
One way is to subtract a monthly minimum (MM) background. 
The monthly minimum approximates the instrumental scatter by {bf calculating} the lowest value of each pixel in all images during a period of about one month. 
However, this method tends to overestimate the scattered light in the streamer belt (equatorial region). 
The lowest value of these pixels during a month usually contains both the scattered light and the steady-intensity value from the corona. 
Hence, the subtraction of MM background 
would output an electron density in streamer regions that is lower than the actual density. 

Another way to account for the scattered light is to subtract a roll minimum (RM) background. 
The roll minimum background is the lowest value of each pixel obtained during a roll maneuver of the spacecraft (instrument) around its optical axis. 
Because the coronal polar regions are much darker than the equatorial ones, the lowest pixel values in the equatorial region 
during the roll maneuver are nearer to the value of the scattered-light intensity than the MM. 

The sensitivity of the COR1-B instrument decreases at a rate of about $0.25$~\% per month \citep{Thompson_2008}. 
Moreover, variations in the spacecraft's distance from the Sun cause changes of the amount of scattered light in the coronagraph images. 
But the roll manoeuvres are done rarely. 
Therefore, it is impossible to use an RM background obtained in one month for data for the following month 
when the highest possible photometric accuracy is needed. A background image for the period between the roll maneuveres can be derived from interpolation of the RM backgrounds over time in a such a way that this temporal dependence follows 
the temporal dependence of the MM backgrounds, because the MM background images are available for every month. 
In our study we used the approach implemented by W. Thompson in the SolarSoft IDL routine {\sf secchi\_prep} with the keyword parameter {\sf calroll}. 

The photometric calibration is based on Jupiter's passage through the COR1 FOV  \citep{Thompson_2008}. 

After subtracting the scattered light, a median filter with a width of three pixels was applied 
to reduce anomalously bright pixels caused by cosmic rays. 
Then, for the CR 2066 reconstruction, every third image pixel was selected (resulting in a $340\times 340$ pixel image from original $1024\times 1024$) to reduce the computer memory size. 
In case of CR 2112 and 2131 reconstructions, we used original $512\times 512$ input data resolution. 
We used the reconstruction domain as a spherical grid with a size of $50\times180\times360$ covering 
heliocentric distances from $1.5$ to $4\ \mathrm{R}_\odot$, Carrington latitudes from $-90$ to $90^\circ$, 
and Carrington longitudes from $0$ to $360^\circ$, respectively.

The inversion was performed for the function
\begin{equation}
F=\left | \mathbf{A}\cdot\mathbf{X}-\mathbf{Y} \right |^2 + \mu \left | \mathbf{R}\cdot\mathbf{X} \right |^2. 
\label{MinFun}
\end{equation}
Here, the elements $x_j$ of the column matrix $\mathbf{X}$ contain the values of electron density [$N_e$] in the grid cells 
with index $j=1,...,n$, 
and $y_i$ is the data value for the $i$-th ray, 
where index $i=1,...,m$ accounts for both the viewing direction and pixel position in the image.  
The element $a_{ij}$ of the matrix $\mathbf{A}$ represents the intersection of 
volume element $j$ with LOS related to pixel $i$, multiplied by the kernel function 
that is defined by the Thompson scattering effect for the pB-intensity signal (see Equation (\ref{ThompsonScat_pB})). 
The second term on the right-hand side of Equation (\ref{MinFun}) is the 
regularization term that minimizes the effects of noise and data gaps \citep{Tikhonov_1963}. 
The matrix $\mathbf{R}$ is a diagonal-like matrix such that the regularization is the first-order smoothing term, 
{\it i.e.} operation $|\mathbf{R}\cdot\mathbf{X}|^2$ produces 
the square difference in value between two neighboring grid cells, summed over all cells. 
The regularization parameter [$\mu$] regulates balance between the smoothness of the solution on one hand 
and the noise and reconstruction artifacts on the other.  
The result of the inversion depends on a number of factors, including the number of iterations and the value of $\mu$.
The value of $\mu$ was chosen using the cross-validation method \citep{Frazin_2002}.  
The iterations are performed until the first term in Equation (\ref{MinFun}) became slightly lower than the data noise level, 
which is essentially the Poisson noise in the data. 

The coronal electron density drops very rapidly with distance from the Sun, introducing a wide dynamic range in the data, 
which causes linear artifacts in the reconstruction. 
To increase the contribution of signals from those LOS that pass through the low density regions, 
and to reduce the artifacts in the numerical reconstruction at larger distances from the Sun, 
we applied a set of weighting coefficients (or preconditioning) 
\begin{equation}
w_i =
\frac{1}{\left( y_i^{(FT1)} \right)^2}
\label{WeightingFunGeneral}
\end{equation}
were applied for the first term in Equation (\ref{MinFun}) in such way that $\sum\limits_{j} (w_i a_{i,j} x_j) = w_i y_i $. 
Here, $y_i^{(FT1)}$ is the inverse Fourier transform of the function $y_i(r_p,\phi_p)$ on $\phi_p$
with harmonics taken up to first order, where $y_i(r_p,\phi_p)$ is the data value at the position $(r_p,\phi_p)$ 
in the polar coordinate system for some particular image.  
The value of $r_p$ is fixed for a given pixel and set equal to the radial distance from the center of the Sun's disk to the pixel.  
A more detailed description of used tomography method is given in \cite{Kramar_2009}. 
The error estimation of the tomographic method has been investigated in \cite{Kramar_2014}. 
The reconstruction results are discussed in Section \ref{Sect_CR2131}.

\subsection{Tomography for Emissivity from STEREO/EUVI Data}
\label{Sect_EUVI_Tomo}

The STEREO/EUVI instrument observes the solar corona at heights up to about $1.7 \ \mathrm{R}_\odot$ 
in four spectral channels (171, 195, 284, and 304 \AA) that span the 0.1 to 20 MK temperature range 
\citep{Wuelser_2004, Howard_2008}. 
The measured coronal emission in the 171, 195, and 284 \AA \  channels can be represented as the result of emission integrated over the LOS as 
\begin{equation}
I(\hat{\vec{e}}_\textrm{LOS},\vec{\rho})=
k  \int \limits_\textrm{LOS} \varepsilon(\vec{r}) \mathrm{d} \ell ,
\label{EUVI_Emiss_LOS}
\end{equation}
where $\varepsilon(\vec{r})$ is the emissivity at the position $\vec{r}$ in the selected channel, 
{\it i.e.} light intensity (in photons per second for example) emitted per unit volume, per unit solid angle. 
The coefficient $k$ accounts for pixel size, aperture, and distance to the Sun. 

As input, we used EUVI 195 \AA \  images calibrated by applying IDL SolarSoft routines. 
To reduce anomalously bright pixels caused by cosmic rays, the IDL SolarSoft routine {\sf despike\_gen} was applied. 
Three images taken with about two hours difference were averaged into one. 
Three averaged images per day were taken during a period of half a solar rotation. 
Then, every fourth pixel was taken, resulting in $512\times512$ input image. 

\begin{figure}[!t]
\includegraphics*[bb=63 298 553 620,width=0.9\linewidth]{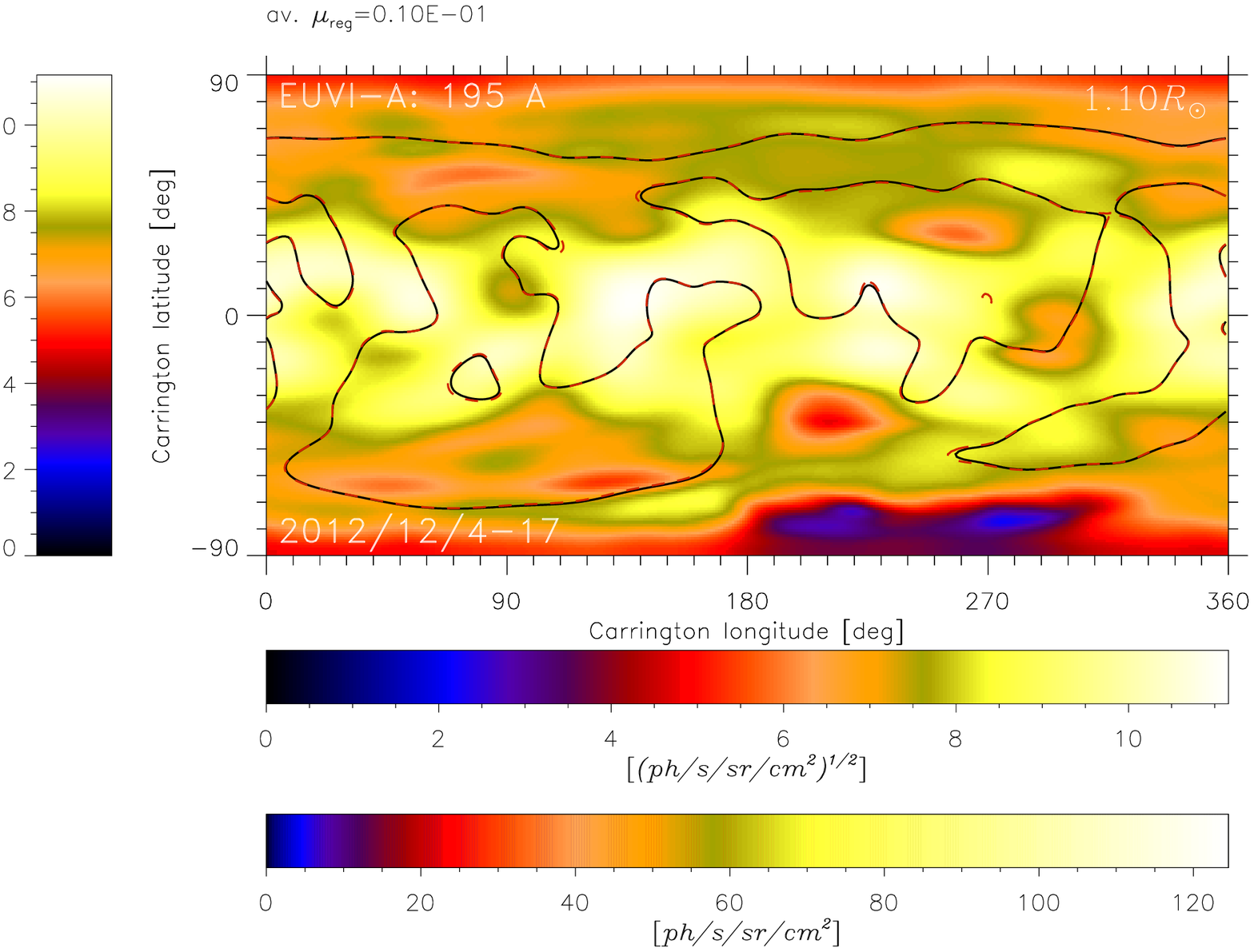}
\caption{Spherical cross-section of the reconstructed 3D EUVI 195 \AA \  emissivity in square root scale at heliocentric distances of $1.1 \ \mathrm{R}_\odot$. 
The reconstruction is obtained by tomography based on COR-1 data obtained during December 4-17, 2012 (CR 2131). 
Solid black and dashed red lines indicate the magnetic neutral line in PFSS model 
with source surface located at $1.5$ and $2 \ \mathrm{R}_\odot$, respectively.}
\label{Fig_CR2131_EUVI_Rec_sph}
\end{figure}

We inverted $\varepsilon(\vec{r})$ in the same manner as for the electron density 
in the white-light tomography with $K$ and $N_e$ in Equation (\ref{ThompsonScat_pB}) substituted by $k$ and $\varepsilon$, respectively, 
according to Equation (\ref{EUVI_Emiss_LOS}). 
The inversion result is 
the 3D emissivity distribution for the EUVI 195 \AA \  channel in the coronal range from $1.05$ to $1.5\ \mathrm{R}_\odot$. 
Figure \ref{Fig_CR2131_EUVI_Rec_sph} shows a spherical cross-section of the reconstructed EUVI 195 \AA \  emissivity 
at a heliocentric distance of $1.1 \ \mathrm{R}_\odot$ for CR 2131. 
The reconstruction results are discussed in Section \ref{Sect_CR2131}.

\section{3D Coronal Structure during CR 2131}
\label{Sect_CR2131}

\begin{figure}[!t]
\includegraphics*[bb=63 298 553 620,width=0.9\linewidth]{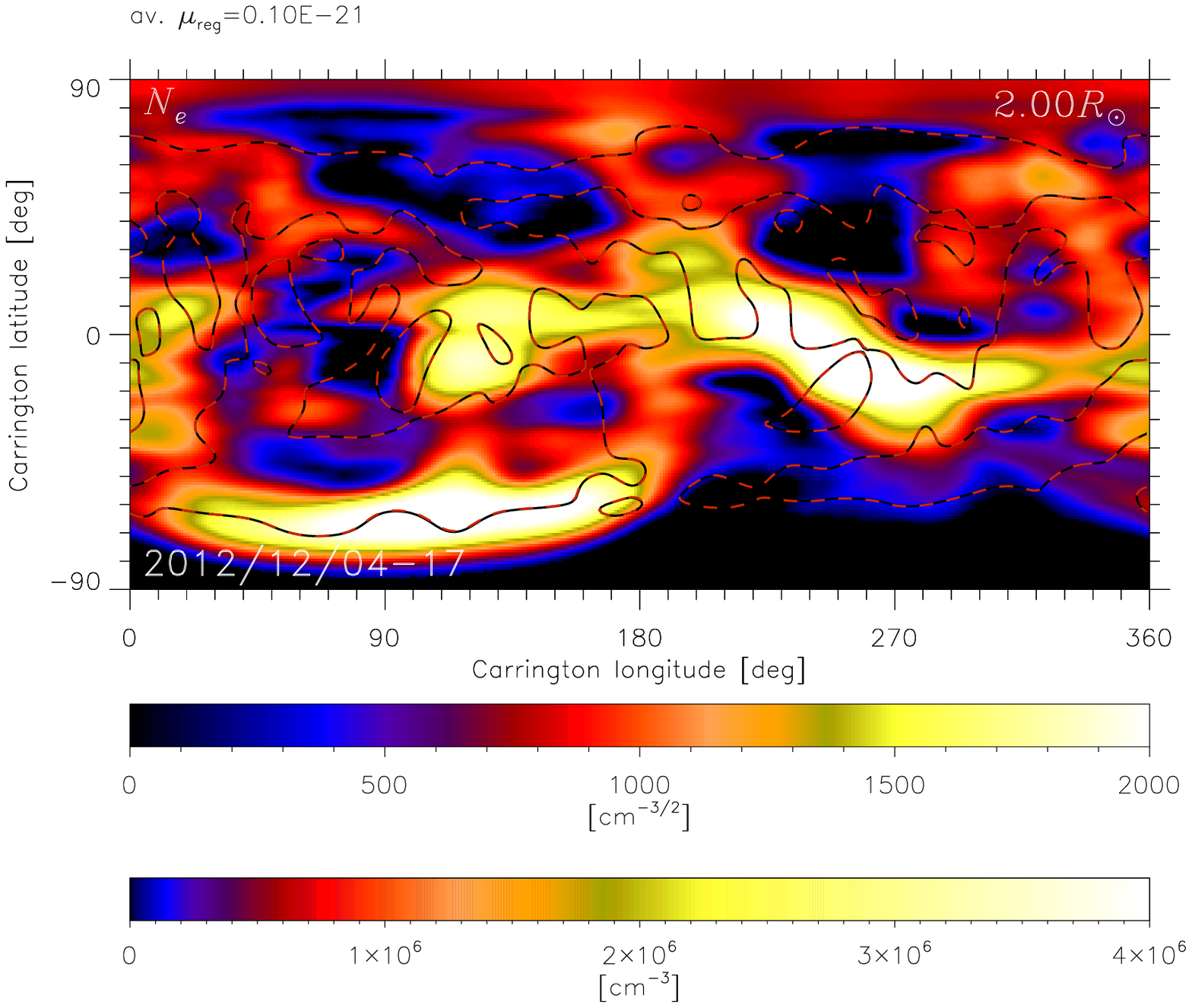}
\caption{Spherical cross-section of the reconstructed electron density in square root scale 
at heliocentric distance of $2 \ \mathrm{R}_\odot$. 
The reconstruction is obtained by tomography based on COR-1 data obtained during December 4-17, 2012 (CR 2131). 
Solid black and dashed red lines indicate the magnetic neutral line in PFSS model 
with source surface located at $1.5$ and $2 \ \mathrm{R}_\odot$, respectively.}
\label{Fig_CR2131_Rec_sph}
\end{figure}

CR 2131 represents the period near maximum of solar activity cycle. 
In this study, we performed two types of tomographic reconstructions: 
3D reconstruction of the electron density based on STEREO/COR1 data 
and 3D reconstruction for the EUVI 195 \AA \  emissivity 
[units of photons s$^{-1}$ sr$^{-1}$ cm$^{-3}$] based on STEREO/EUVI data. 
In order to demonstrate a general structure of the coronal streamer belt for CR 2131, 
Figure \ref{Fig_CR2131_Rec_sph} shows a spherical cross-section of the electron density at 
the heliocentric distance of $2 \ \mathrm{R}_\odot$ 
and Figure \ref{Fig_CR2131_EUVI_Rec_sph} shows a spherical cross-section of 
the EUVI 195 \AA \  emissivity at the heliocentric distance of $1.1 \ \mathrm{R}_\odot$. 

\begin{figure}[ptbh]
\includegraphics*[bb=75 345 543 718,width=0.46\linewidth]{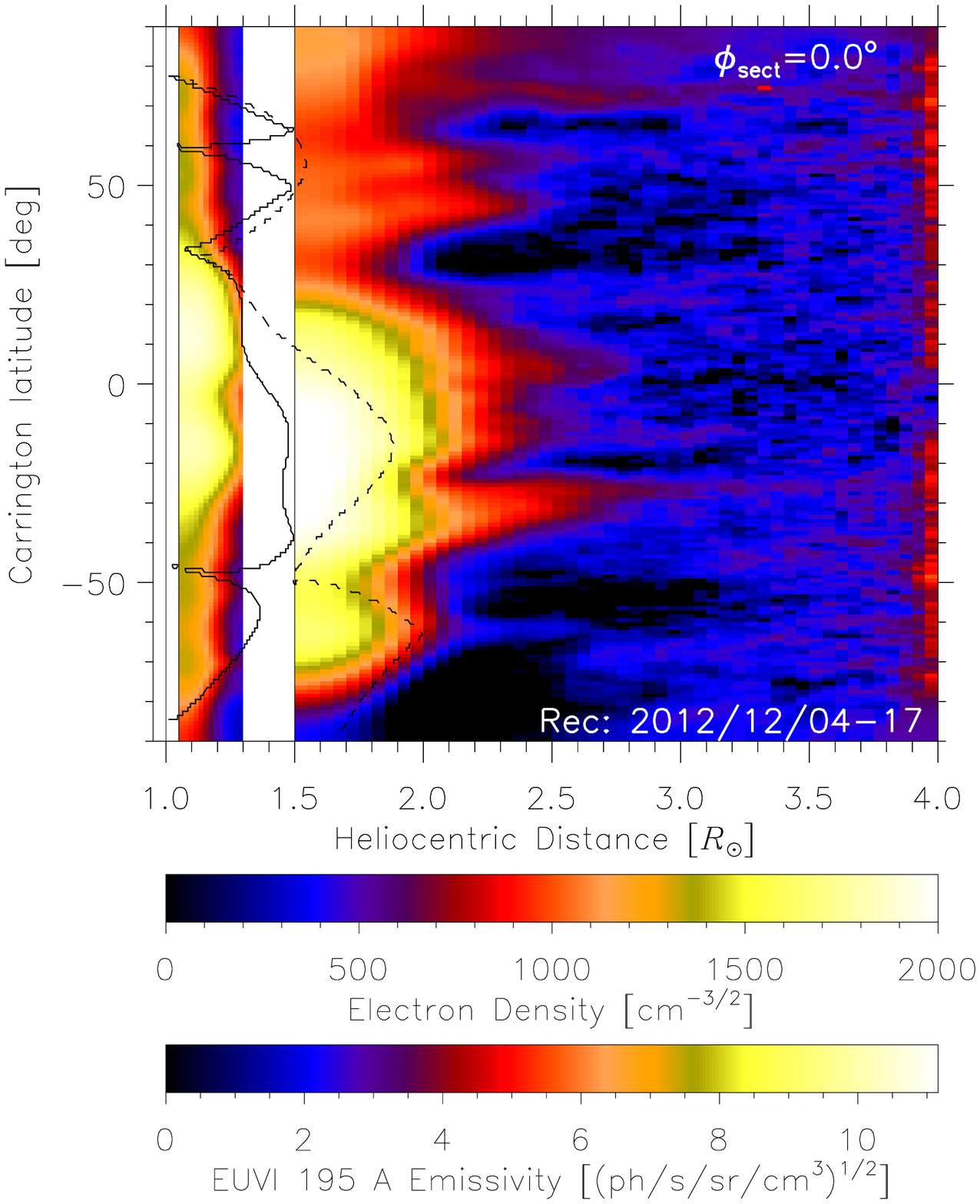}
\includegraphics*[bb=75 345 543 718,width=0.46\linewidth]{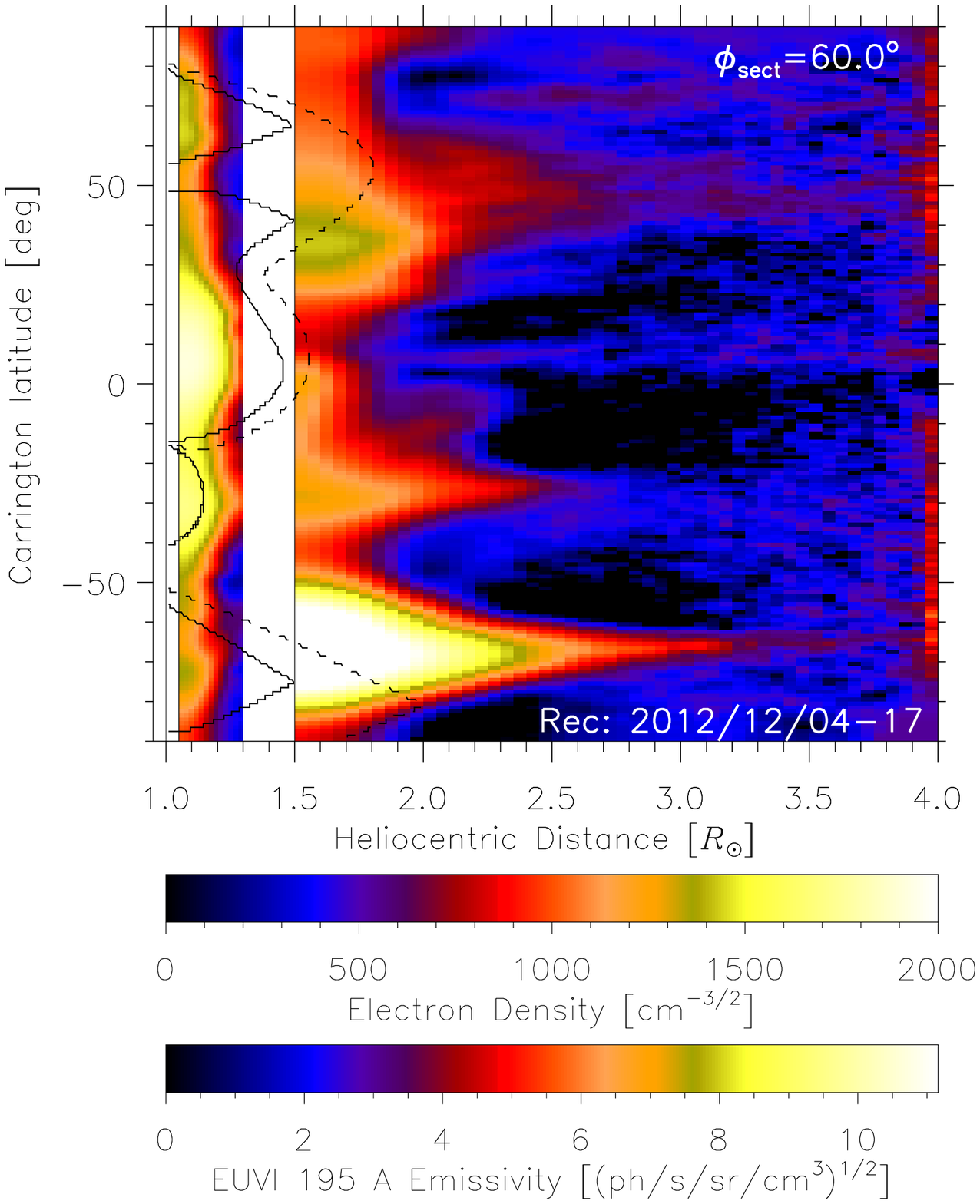}\\
\includegraphics*[bb=75 345 543 718,width=0.46\linewidth]{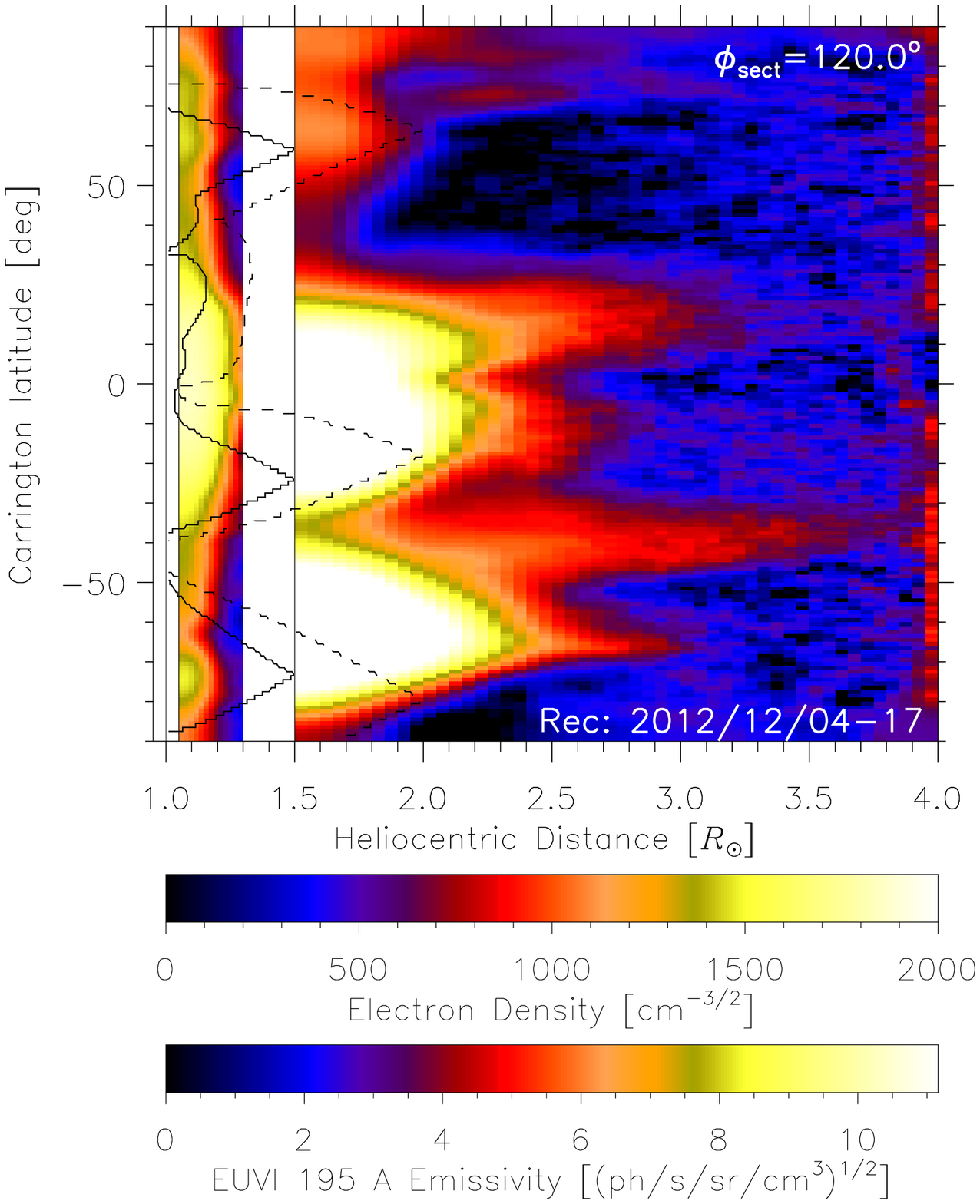}
\includegraphics*[bb=75 345 543 718,width=0.46\linewidth]{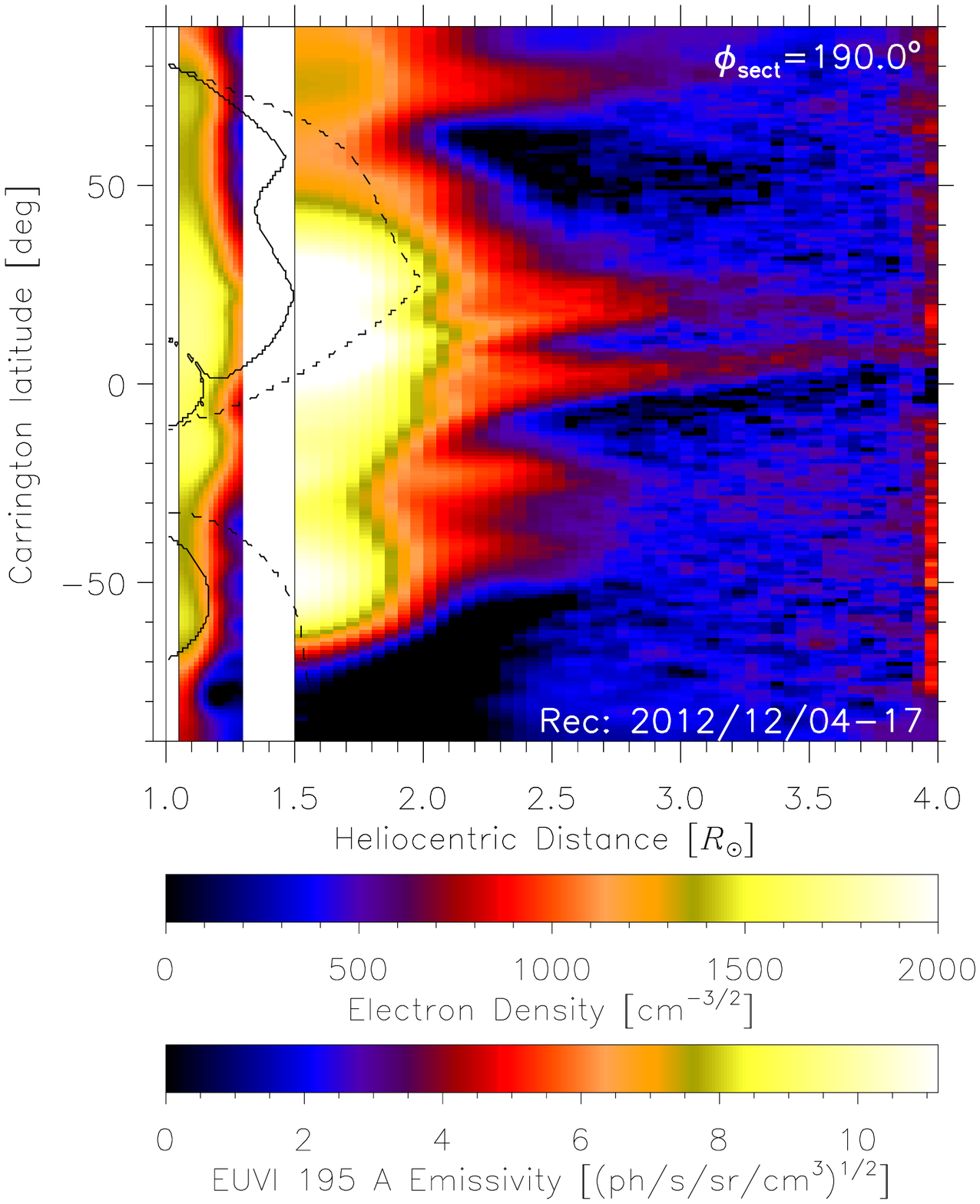}\\
\includegraphics*[bb=75 301 543 718,width=0.46\linewidth]{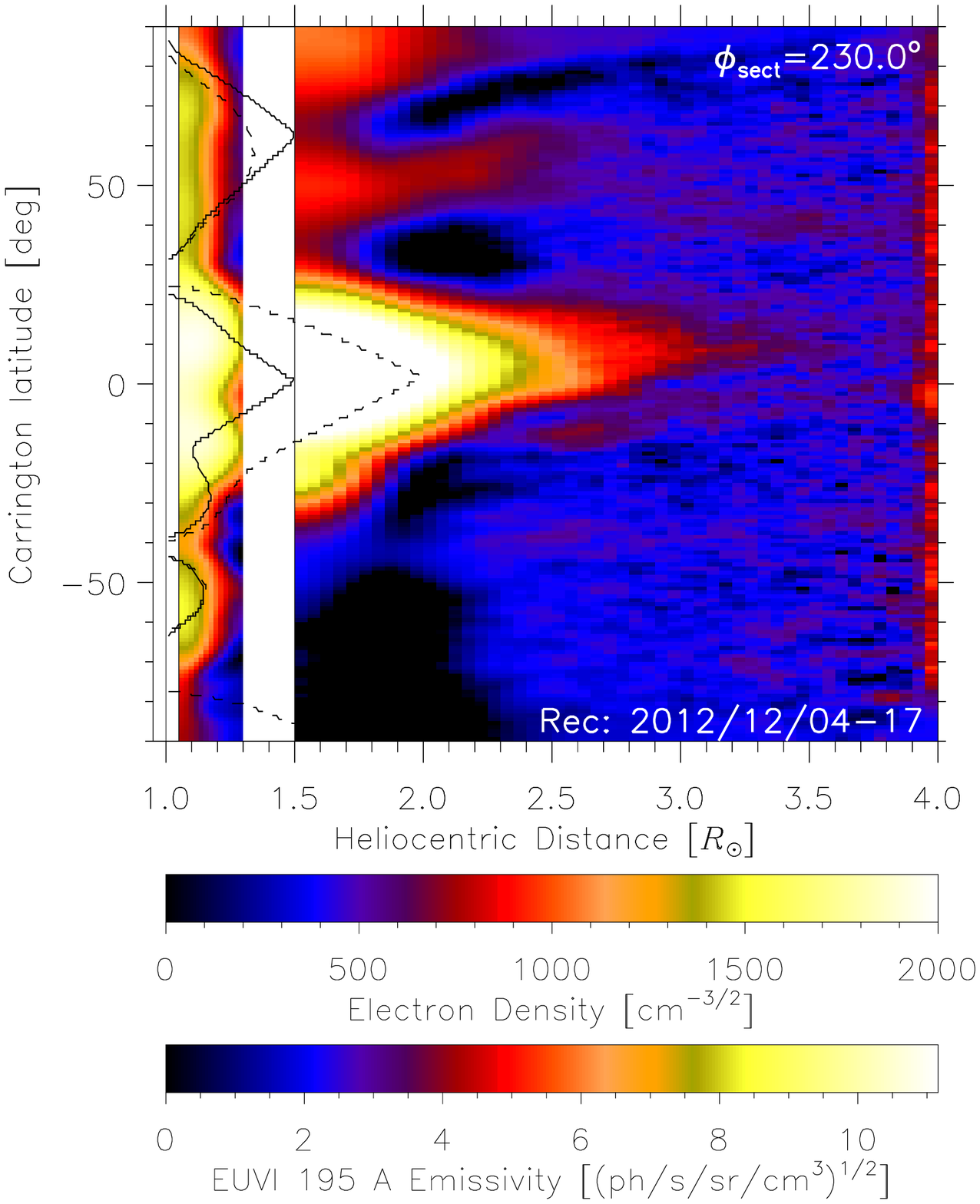}
\includegraphics*[bb=75 301 543 718,width=0.46\linewidth]{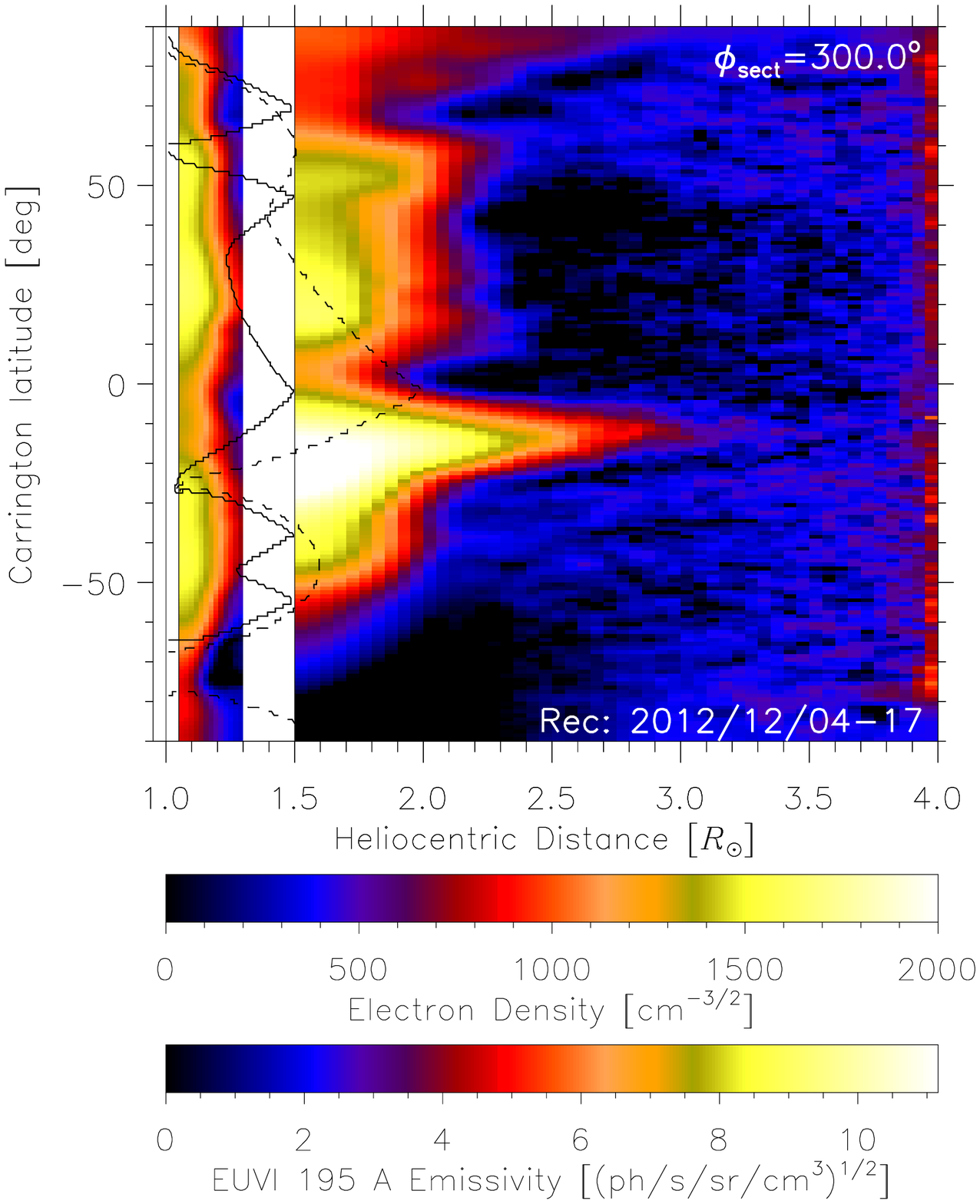}\\
\includegraphics*[bb=75 136 543 215,width=0.46\linewidth]{f3a.ps}
\includegraphics*[bb=75 219 543 298,width=0.46\linewidth]{f3a.ps}
\caption{Reconstructions for CR 2131 based on COR1 data 
(electron density in the range from $1.5$ to $4 \ \mathrm{R}_\odot$) 
and EUVI 195 \AA \  data (emissivity in the range from $1.05$ to $1.29 \ \mathrm{R}_\odot$). 
Carrington longitudes for cross-sections are shown at upper right corners. 
The figure with a set of all cross-sections is available in the Electronic Supplementary Material. 
The contour black lines show boundaries between open and closed magnetic-field structures for 
the PFSS models with Source Surface located at $1.5$ and $2.0 \ \mathrm{R}_\odot$.}
\label{Fig_RSS_COR1_phi_CR2131}
\end{figure}

Figure \ref{Fig_RSS_COR1_phi_CR2131} shows several meridional cross-sections of the electron density 
(range from $1.5$ to $4 \ \mathrm{R}_\odot$) 
and EUVI 195 emissivity (range from $1.05$ to $1.29 \ \mathrm{R}_\odot$). 
The figure with a set of all cross-sections is available in the Electronic Suplemental Material. 

The black contour lines in Figure \ref{Fig_RSS_COR1_phi_CR2131} show boundaries between open and closed 
magnetic-field structures in two PFSS models with Source Surface heliocentric distances 
[$R_\textrm{ss}$] at $1.5$ and $2.0 \ \mathrm{R}_\odot$. 
The PFSS model with $R_\textrm{ss}=2.0 \ \mathrm{R}_\odot$ does not coincide with the derived positions of the 
streamer and pseudo-streamer as well as with the coronal hole positions indicated by 
the STEREO/EUVI 195 \AA \  emissivity 3D reconstruction. 
The PFSS model with $R_\textrm{ss}=1.5 \ \mathrm{R}_\odot$ appears to fit the latter structures better. 

\begin{figure}[ptbh]
\includegraphics*[bb=75 345 543 718,width=0.46\linewidth]{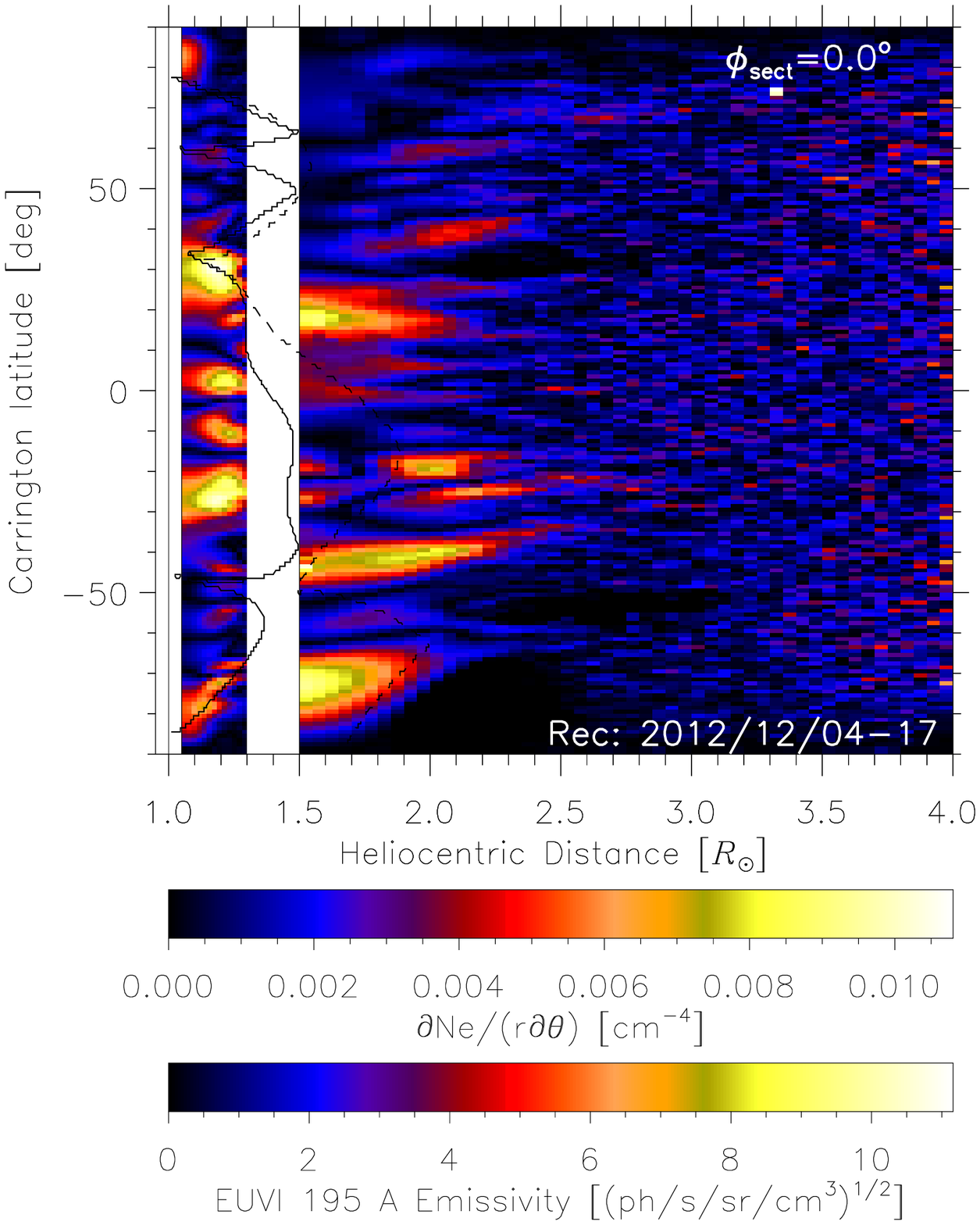}
\includegraphics*[bb=75 345 543 718,width=0.46\linewidth]{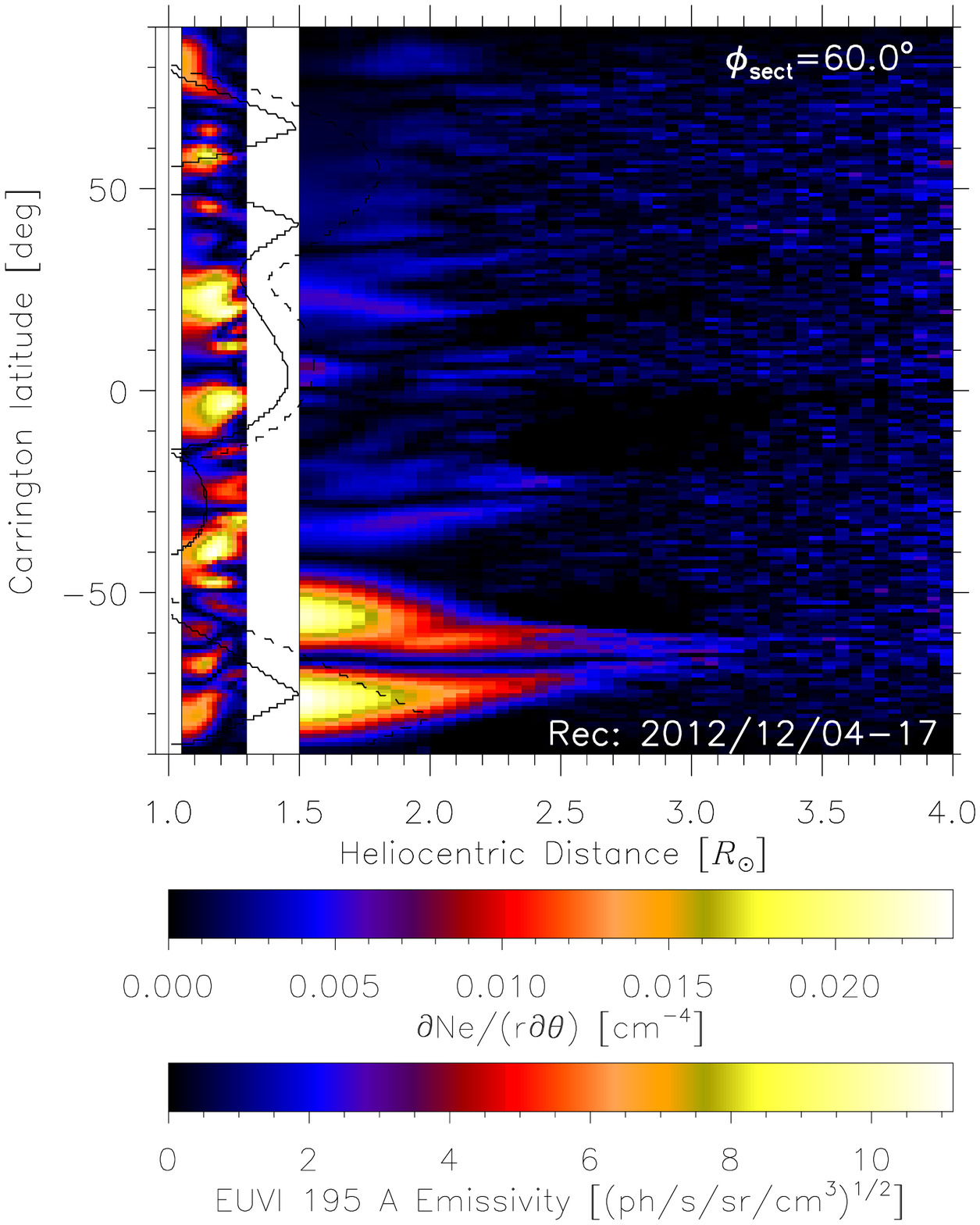}\\
\includegraphics*[bb=75 345 543 718,width=0.46\linewidth]{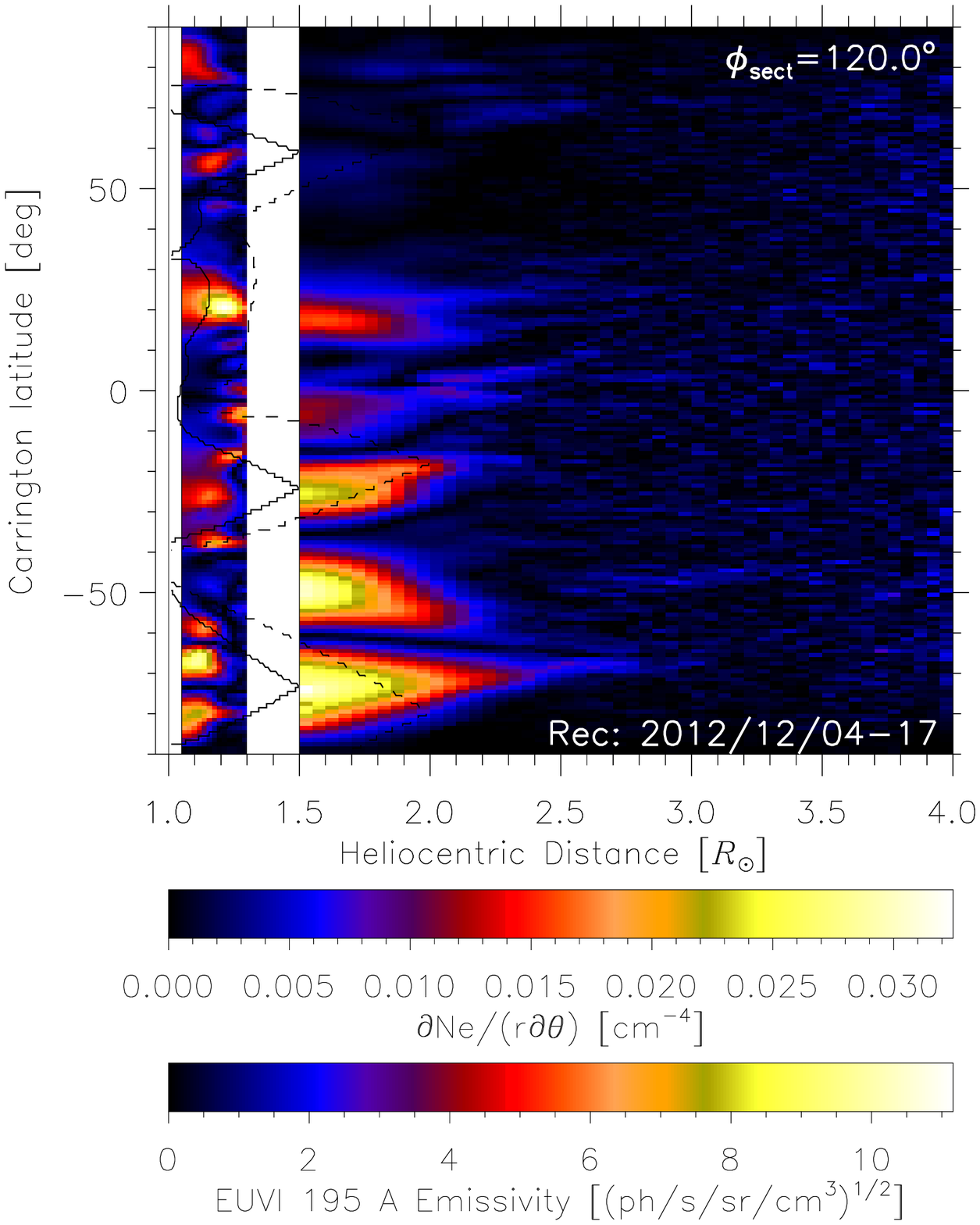}
\includegraphics*[bb=75 345 543 718,width=0.46\linewidth]{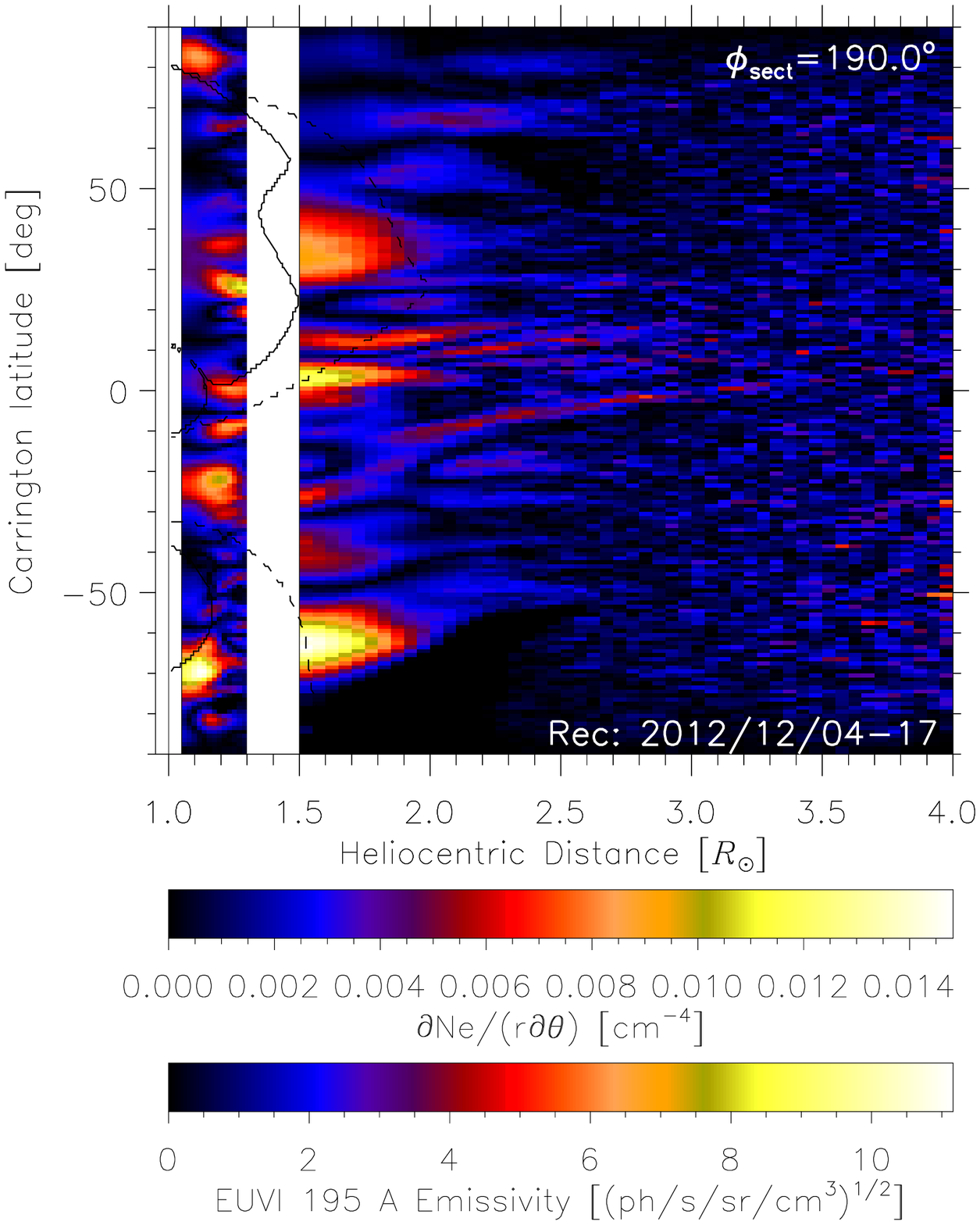}\\
\includegraphics*[bb=75 301 543 718,width=0.46\linewidth]{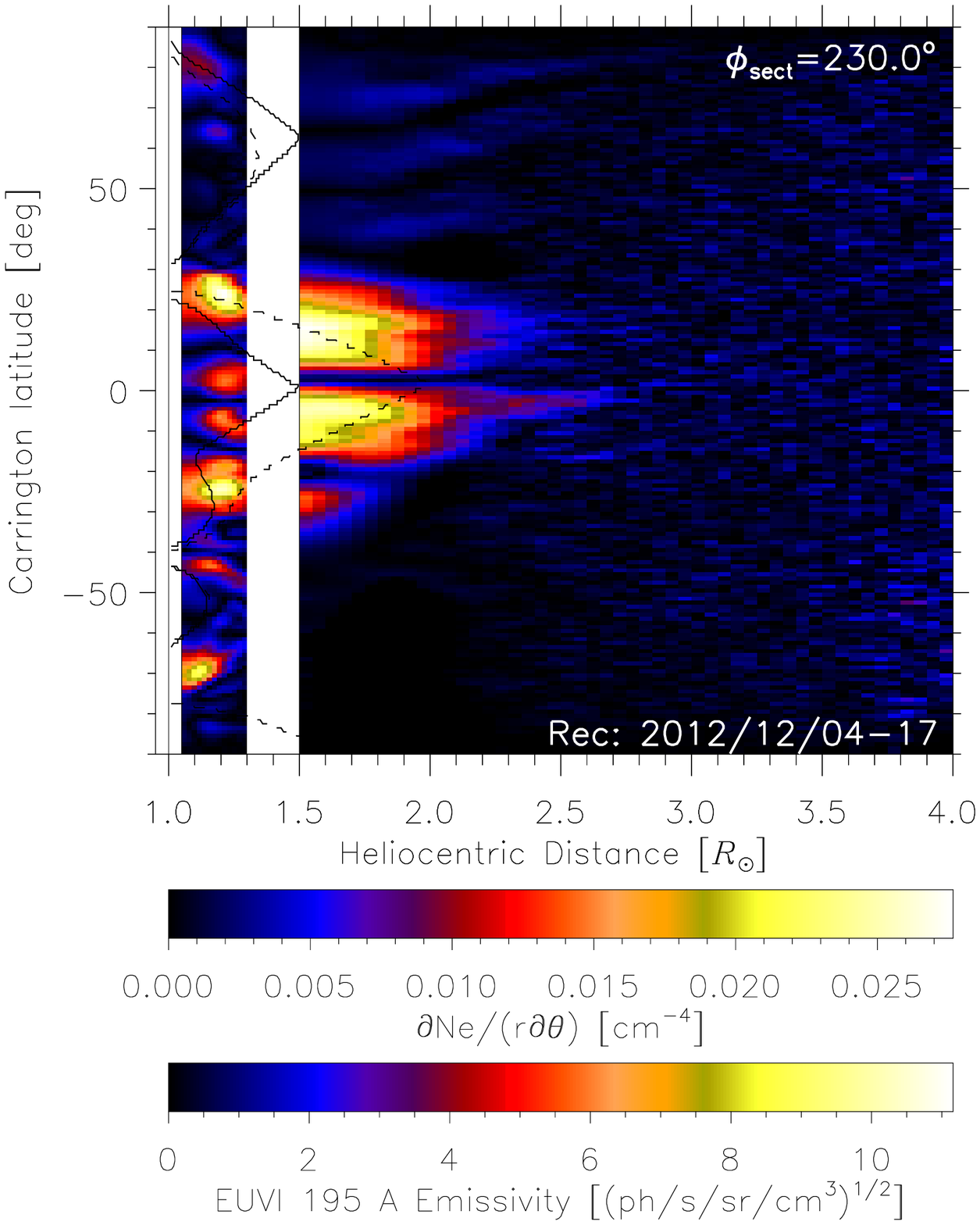}
\includegraphics*[bb=75 301 543 718,width=0.46\linewidth]{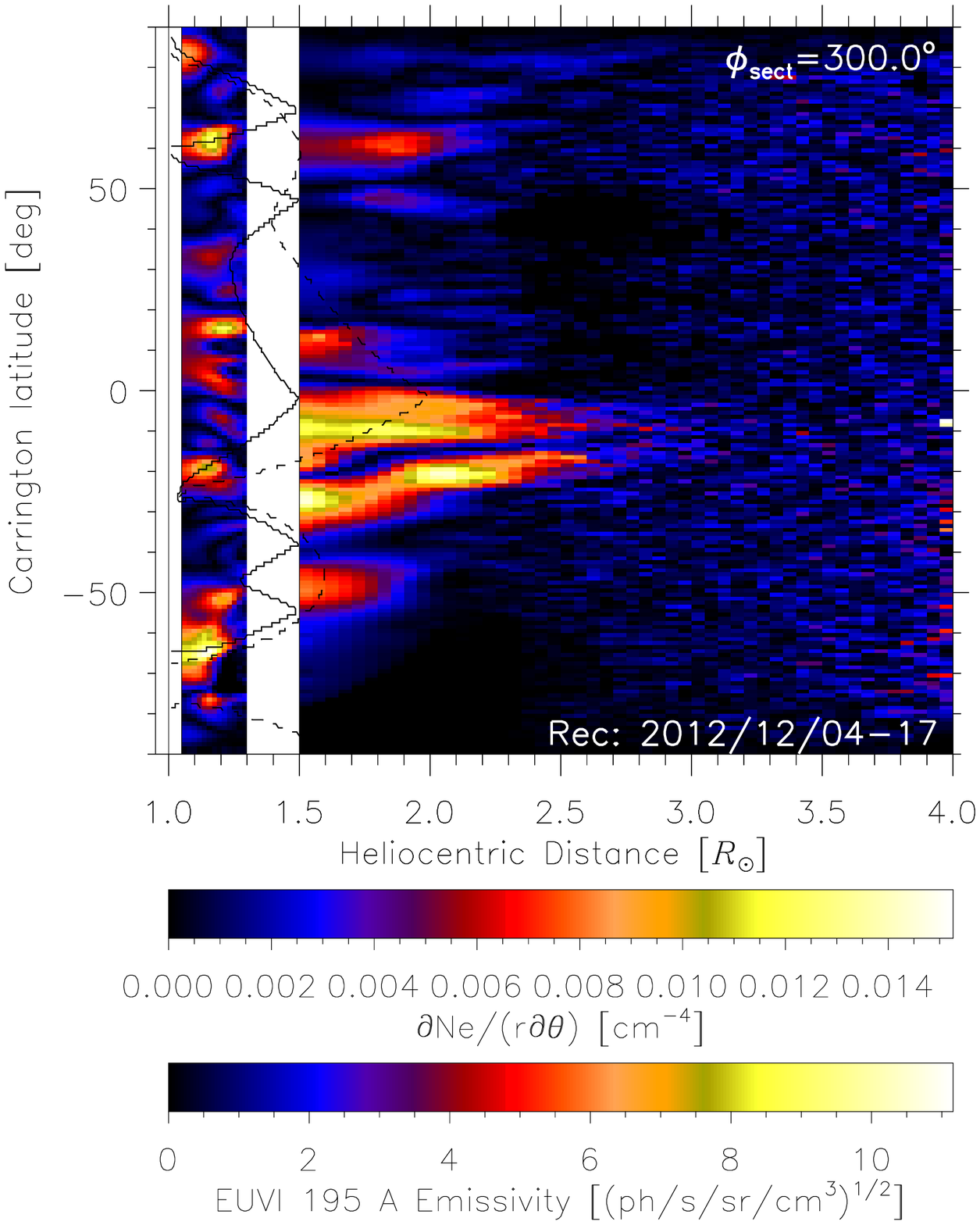}\\
\caption{Longitudinal cross-sections of gradient of electron density and 
EUVI 195 emissivity, 
$\partial N_e/\partial\theta$ and $\partial \varepsilon_{195}/\partial\theta$, for CR 2131. 
Carrington longitudes for cross-sections are shown at upper right corners. 
Color scales are different for different cross-sections to enhance the images. 
The contour black lines show boundaries between open and closed magnetic-field structures for 
the PFSS models with Source Surface located at $1.5$ and $2.0 \ \mathrm{R}_\odot$.}
\label{Fig_RSS_COR1_phi_CR2131_grad}
\end{figure}

It was previously demonstrated that the position of maximal gradient of the density and 
195 \AA \ emissivity can be an indicator for the position of boundary between 
closed and open magnetic structures \cite{Kramar_2014}. 
Therefore, in order to exhibit the possible real boundary positions, 
Figure \ref{Fig_RSS_COR1_phi_CR2131_grad} shows 
gradient of electron density and EUVI 195 emissivity, 
$\partial N_e/\partial\theta$ and $\partial \varepsilon_{195}/\partial\theta$, 
for CR 2131 and for the same longitudinal cross-sections as in 
Figure \ref{Fig_RSS_COR1_phi_CR2131}. 
Both Figures \ref{Fig_RSS_COR1_phi_CR2131} and \ref{Fig_RSS_COR1_phi_CR2131_grad} 
demonstrate that the magnetic field configuration during CR 2131 has a tendency 
to become radially open at heliocentric distances below $\sim 2.5 \ \mathrm{R}_\odot$. 
This is lower than corresponding distance during the deep solar minimum near CR 2066 \citep{Kramar_2014}.

\begin{figure}[!t]
\includegraphics*[bb=92 370 543 715,width=0.49\linewidth]{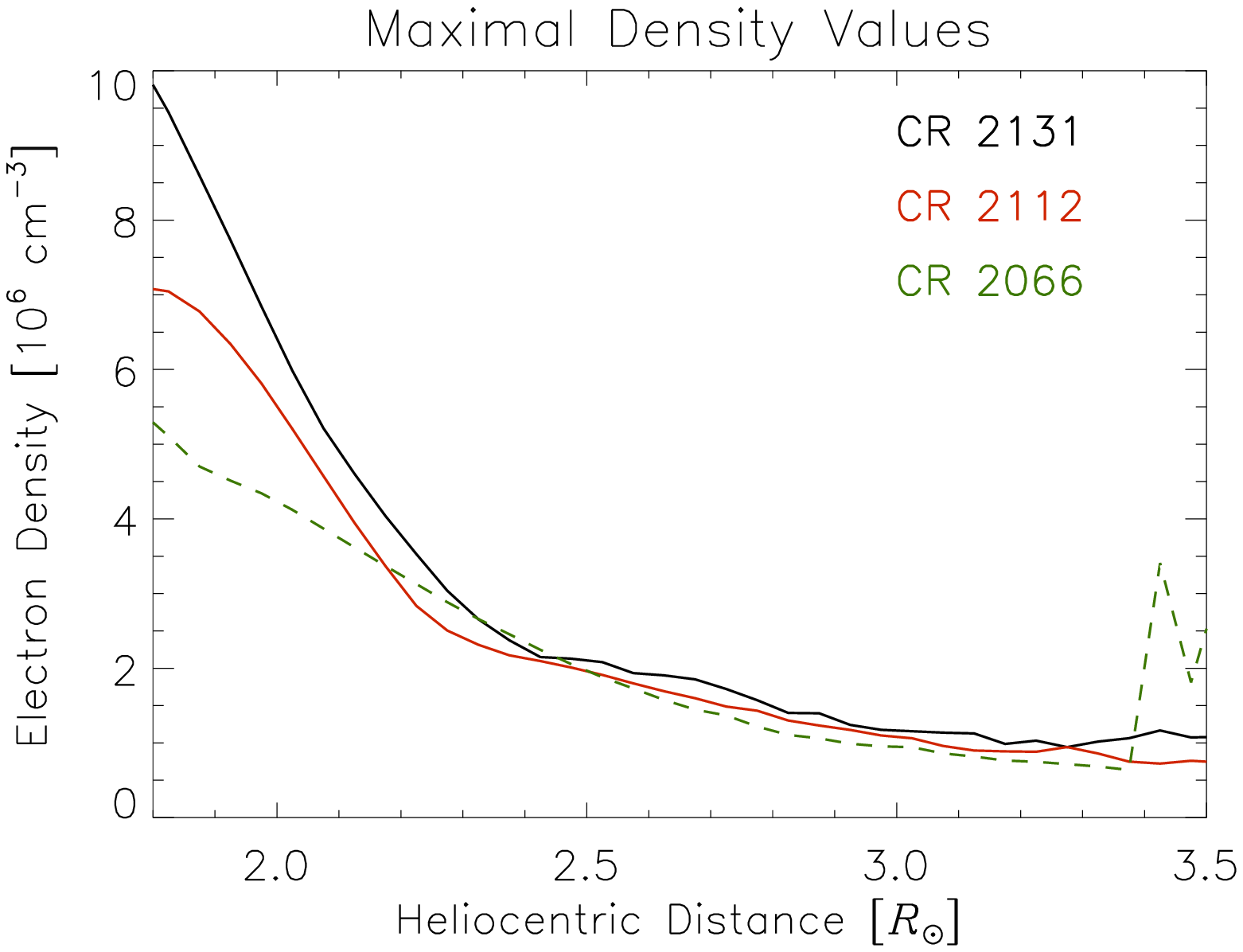}
\includegraphics*[bb=92 370 543 715,width=0.49\linewidth]{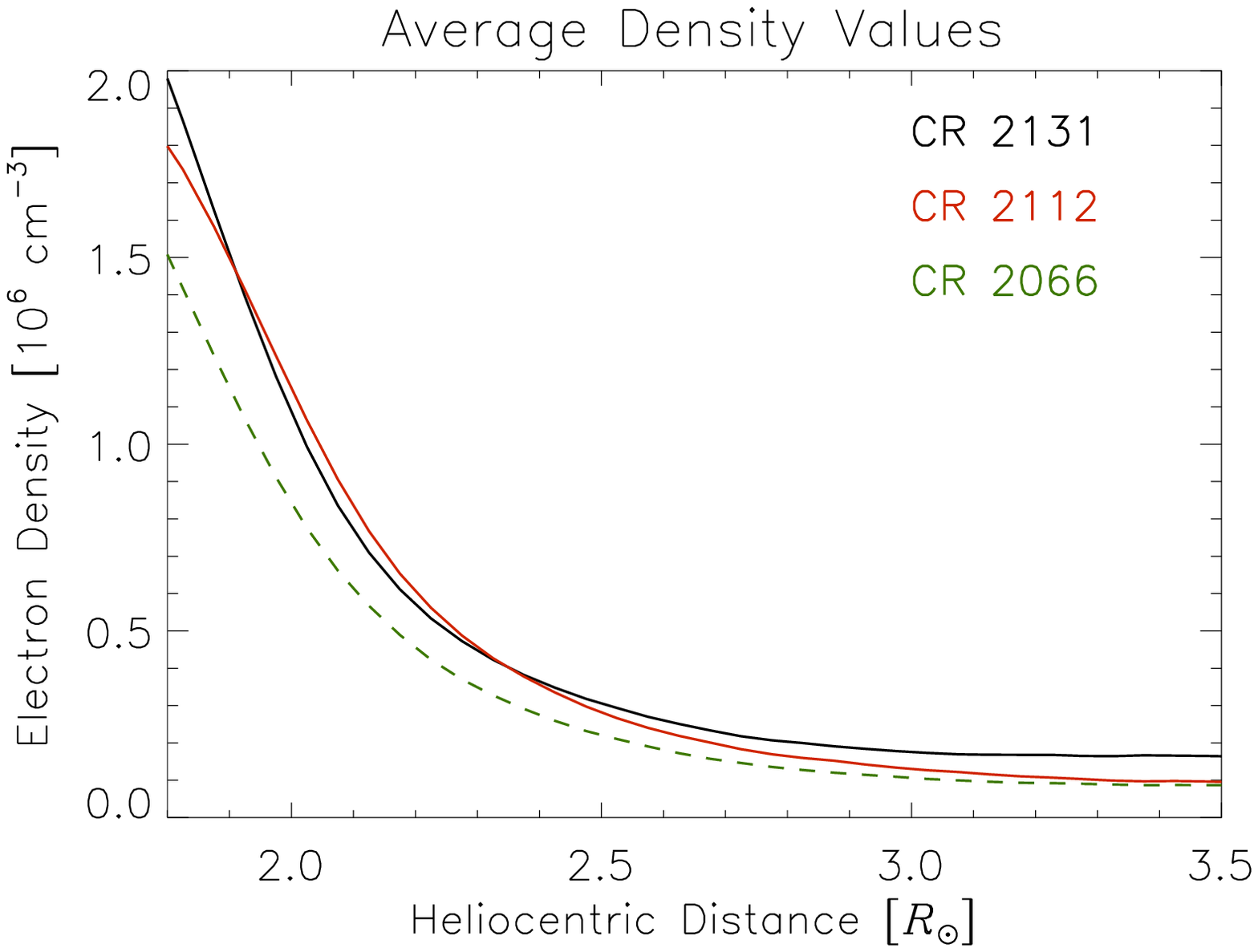}
\caption{Maximal (left panel) and average over a spherical surface area (right panel) electron density values at 
different heliocentric distances for CR 2066 (dashed green), 
2112 (solid red), and 2131 (solid black). 
CR 2066 corresponds to deep solar minimum, 
while CR 2131 corresponds to period near solar maximum.}
\label{Fig_Rec_Comp}
\end{figure}

Figure \ref{Fig_Rec_Comp} shows the radial profiles of the maximum (left panel) and average (right panel) electron density values  at different heliocentric distances 
for CR 2066 (dashed green), 2112 (solid red), and 2131 (solid black). 
CR 2066 corresponds to deep solar minimum, while CR 2131 corresponds to period near solar maximum. 
The maximum values of coronal density mostly reflect the closed magnetic field structures represented by solar active regions, while average values include contributions from both open and closed field regions. The plot suggests that the maximum coronal density at the solar maximum is larger than the coronal density at two Carrington rotations representing the periods of lower magnetic activity. Specifically, at the lower boundary (at 1.8 $R_\odot$) of the reconstruction region the maximum density at solar  maximum is more that twice greater than the maximum coronal density at solar minimum represented by CR 2066. This reflects the fact that the total unsigned magnetic flux that is structured in the form of coronal active regions during solar maximum  is a factor of 2-3 greater than the emerging magnetic flux during solar minimum \citep{Solanki_2002AA}. The average coronal density is a factor of 1.6 greater during solar maximum as compared to solar minimum, because it reflects the contribution of plasma emission formed in the diffuse corona as well as coronal active regions.

\begin{figure}[!t]
\includegraphics*[width=\linewidth]{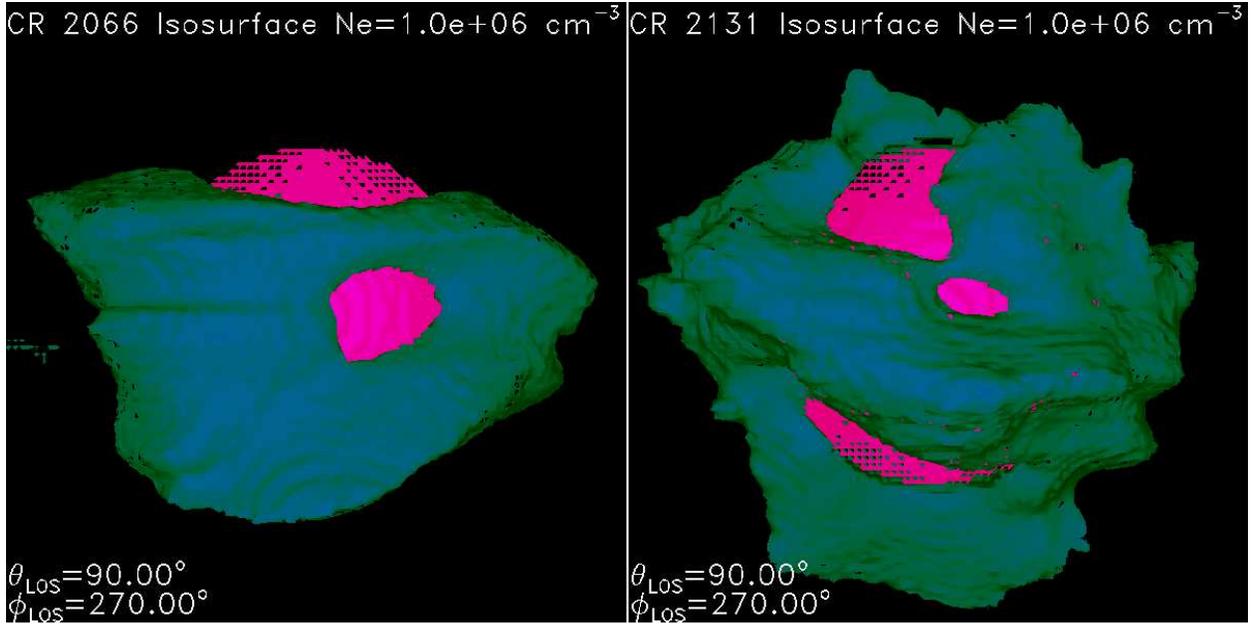}
\caption{Isosurface of coronal electron density value of $10^6\textrm{cm}^{-3}$ (in blue) 
for CR 2066 (left) and 2131 (right). 
The orange sphere is located at $1.5 R_\odot$ just for scaling purpose. 
The movie demonstrating the isosurfaces for the whole longitudinal projection range is provided in 
online version of the paper.}
\label{Fig_Rec_Comp_3D}
\end{figure}

Figure \ref{Fig_Rec_Comp_3D} shows the snapshot of the isosurface of coronal electron density value of $10^6\textrm{cm}^{-3}$ (in blue) 
for CR 2066 occured at solar minimum (left) and CR 2131 at solar maxium (right). The corresponding movie showing the density isosurface over these two Carrington rotations are available provided in online version of the paper. The figure and the movie clearly demonstrate that the filling factor of the plasma with the density of  $10^6\textrm{cm}^{-3}$ increases at solar maximum, which is consistent with Figure \ref{Fig_Rec_Comp}.

\section{Conclusion and Outlook}

We applied the tomography method to STEREO-B/COR1 data to derive the reconstructions of the 3D global coronal electron density and EUVI 195 \AA \ emissivity based on STEREO/ COR1 and STEREO/EUVI observations near solar maximum represented by Carrington rotation CR 2131 (December 2012).  We complemented the tomography with MHD simulations to obtain the open/closed magnetic field boundaries in the solar corona. Because the 3D reconstructions are based entirely on {\it coronal} data, these results could serve as an independent test and/or as an additional constraint for the models of the solar corona. Specifically, we used the PFSS model for different source surface distances as a test case for the reconstructed 3D electron density and EUVI 195 emissivity structures. 

	We have shown that magnetic field structures deviate significantly from the PFSS approximation as evident from the derived boundaries of open/closed field for both solar minimum \citep{Kramar_2014} and solar maximum. 
This suggests that potential field approximation for the low corona is not valid approximation when describing its large-scale structures. It is important to note that the magnetic field becomes radial at heights greater than $2.5 R_\odot$ during solar minimum, while its open up at heights below $\sim 2.5 R_\odot$ during solar maximum. 

Our studies also show that the average and the maximum electron densities in the low solar corona at heights 
$\sim 1.8 R_\odot$ are about 2 times greater than that obtained during solar minimum. 
This is an important result that can be extrapolated to coronal densities of active and young solar-like stars. These stars show much greater levels of stellar activity in terms of at $>10$ times greater surface magnetic flux, up to 1000 times more luminous in X-rays and at least 10 times denser and hotter corona. These results open up an interesting opportunity to provide scaling of the electron density of the solar corona with different levels of magnetic activity traced by its X-ray luminosity and average surface magnetic flux. These scaling laws can be then used for characterization of stellar corona of active stars, which may provide insights on the nature of stellar coronal heating at various phases of their evolution.

Applied in this study method can also be used for verification of 
3D global coronal vector magnetic field obtained by 
vector tomography based on coronal polarimetric observations \citep{Kramar_2016}.

\clearpage

\clearpage

\clearpage

\end{document}